\documentclass[10pt,aps,pra,reprint,superscriptaddress,showkeys]{revtex4-1}
\usepackage{graphicx}
\usepackage[colorlinks=true,linkcolor=blue,citecolor=blue,urlcolor=blue]{hyperref}
\usepackage[exponent-product=\cdot,range-phrase = --,range-units = single,per-mode=symbol]{siunitx}
\makeatother
\begin{document}

\title{Fields of View for Environmental Radioactivity}

\author{Alex Malins}\email[Corresponding author: ]{malins.alex@jaea.go.jp}
\affiliation{Center for Computational Science \& e-Systems, Japan Atomic Energy Agency, 178-4-4 Wakashiba, Kashiwa-shi,
Chiba-ken, 277-0871, Japan}

\author{Masahiko Okumura}
\affiliation{Center for Computational Science \& e-Systems, Japan Atomic Energy Agency, 178-4-4 Wakashiba, Kashiwa-shi,
Chiba-ken, 277-0871, Japan}

\author{Masahiko Machida}
\affiliation{Center for Computational Science \& e-Systems, Japan Atomic Energy Agency, 178-4-4 Wakashiba, Kashiwa-shi,
Chiba-ken, 277-0871, Japan}

\author{Hiroshi Takemiya}
\affiliation{Center for Computational Science \& e-Systems, Japan Atomic Energy Agency, 178-4-4 Wakashiba, Kashiwa-shi,
Chiba-ken, 277-0871, Japan}

\author{Kimiaki Saito}
\affiliation{Fukushima Environmental Safety Center, Sector of Fukushima Research and Development, Japan Atomic Energy Agency, 2-2-2 Uchisaiwai-cho, Chiyoda,
Tokyo, 100-8577, Japan}

\date{\today}

\begin{abstract}
The gamma component of air radiation dose rates is a function of the amount and spread of radioactive nuclides in the environment. These radionuclides can be natural or anthropogenic in origin. The field of view describes the area of radionuclides on, or below, the ground that is responsible for determining the air dose rate, and hence correspondingly the external radiation exposure. This work describes Monte Carlo radiation transport calculations for the field of view under a variety of situations. Presented first are results for natural \textsuperscript{40}K and thorium and uranium series radionuclides distributed homogeneously within the ground. Results are then described for atmospheric radioactive caesium fallout, such as from the Fukushima Daiichi Nuclear Power Plant accident. Various stages of fallout evolution are considered through the depth distribution of \textsuperscript{134}Cs and \textsuperscript{137}Cs in soil. The fields of view for the natural radionuclides and radiocaesium are different. This can affect the responses of radiation monitors to these nuclides if the detector is partially shielded from the ground within its field of view. The field of view also sets the maximum reduction in air dose rates that can be achieved through local decontamination or remediation measures. This maximum efficiency can be determined quickly from the data presented here for the air dose rate versus the spatial extent of radioactive source on the ground.
\end{abstract}


\maketitle

\section{Introduction}

Humans are exposed to external gamma radiation from cosmic sources and from radioactivity in the environment on earth. Environmental gamma radiation sources can be natural or man-made in origin. Examples of the former include \textsuperscript{40}K, and thorium and uranium series radionuclides within the ground, and the latter include fallout from atomic weapons testing and civil nuclear accidents. The level of gamma radiation at a particular location can be quantified by measuring an external radiation dose rate. The radiation protection quantity \textit{ambient dose equivalent rate} is commonly used for this purpose~\cite{ICRP74}.

Gamma rays can travel hundreds of metres in air without undergoing interactions. The U.~S.~National Institute of Standards and Technology (NIST) lists a mass attenuation coefficient ($\mu/\rho$) for \SI{1.0}{\mega \electronvolt} photons in dry air of \SI{6.358e-02}{\square \centi \metre \per \gram}~\cite{Hubbell2004}. For air with density \SI{1.2e-03}{\gram \per \cubic \centi \metre}, this equates to a mean free path of \SI{131}{\metre} between interactions. Therefore, gamma rays originating from a wide area of land can contribute to the air dose rate.

The field of view describes the size of the region containing radioactivity whose radiation contributes significantly to the dose rate at a detection point. The field of view of environmental radionuclides, present on or below the ground, can thus be defined as the volume of earth from which a specified fraction of the total gamma radiation intensity contributing to an air dose rate originates from~\cite{Allyson1994}. Previous authors have characterized the field of view for \SI{1}{\metre} air dose rates from natural and anthropogenic radionuclides distributed within soils. Techniques employed have included first principles calculations for photon fluxes~\cite{Allyson1994,Tyler1996} and Monte Carlo simulations of dose rates~\cite{Iwamoto2011} from increasing area sources on the ground. Others have characterized the angular dependence and the field of view of the air kerma~\cite{Saito1991, Saito1995, Saito1998}.

This paper presents Monte Carlo calculations for the field of view for ambient dose equivalent rates from natural radionuclides distributed uniformly in the ground, and radioactive \textsuperscript{134}Cs and \textsuperscript{137}Cs fallout distributed exponentially with depth. Two applications of the fields of view results are presented within: i) for interpreting differences between KURAMA survey results taken in cars and buses, and ii) for undertaking quick calculations for the effectiveness of land remediation methods.

\section{\label{sec:Methods}Methods}

All calculations were undertaken with the Monte Carlo radiation transport code PHITS~\cite{Sato2013}. The simulations modelled the infinite half-space geometry~\cite{ICRU53}. This geometry consists of a plane which separates a lower region of soil (density $\rho_\mathrm{s}=$~\SI{1.6}{\gram \per \cubic \centi \metre}) and an upper region consisting of air ($\rho_\mathrm{a}=$~\SI{1.2e-03}{\gram \per \cubic \centi \metre}). The elemental compositions of the soil and air materials followed ref.~\cite{Eckerman1993}.

\begin{table}
\caption{\label{tab:sources}Radiation sources and their distributions within soil simulated in this work.}
\begin{ruledtabular}
{\begin{tabular}{l l l}
 Source & Depth distribution & Decay emission data ref.\\\hline
 \textsuperscript{40}K & Uniform & ICRP 107~\cite{ICRP107}\\
 \textsuperscript{232}Th series & Uniform & ICRP 107~\cite{ICRP107} \\
 \textsuperscript{235}U series & Uniform & ICRP 107~\cite{ICRP107} \\
 \textsuperscript{238}U series & Uniform & ICRP 107~\cite{ICRP107} \\
 \textsuperscript{134}Cs & Exponential & NuDat2~\cite{NuDat2} \\
 \textsuperscript{137}Cs & Exponential & NuDat2~\cite{NuDat2} \\
\end{tabular}}
\end{ruledtabular}
\end{table}

The natural potassium radioisotope \textsuperscript{40}K, the isotopes in three thorium and uranium radionuclide decay chains, and two radiocaesium isotopes common in nuclear accident fallout were considered as terrestrial gamma radiation sources. The source types, their depth distributions within the ground, and the references used for their x-ray and gamma radiation emission data are listed in Table~\ref{tab:sources}.

\begin{figure}
 \includegraphics{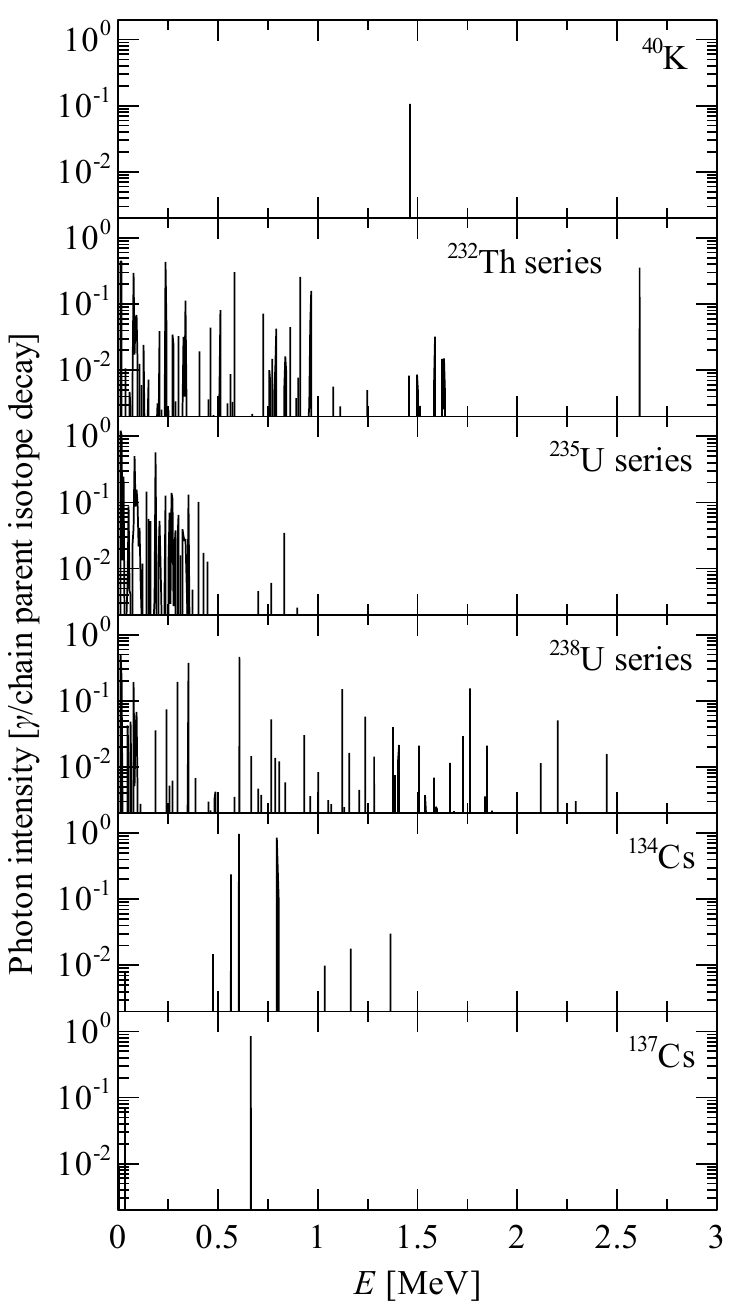}
 \caption{\label{fig:emission_spectra}Photon emission spectra per decay for the different source types. For sources that are a decay chain, the spectra include decay photons from daughter nuclides assumed to be in secular equilibrium and the intensity values shown are applicable per decay of the head isotope in the chain.}
\end{figure}

The photon energy spectra inputted into the simulations for each of the source types are shown in Fig.~\ref{fig:emission_spectra}. All photons with energy greater than \SI{10}{\kilo \electronvolt} and emission probability greater than \SI{0.1}{\percent}, according to the references listed in Table~\ref{tab:sources}, were included in the spectra. Photons were emitted isotropically from the source radionuclides.

The natural radionuclides were modelled with constant activity per unit soil mass at all depths. Caesium was modelled as exponentially distributed with soil depth,
\begin{equation}
A_{\mathrm{m}}(\zeta)=A_{\mathrm{m},0}\exp{(-\zeta/\beta)}\,,
\end{equation}
\noindent where $A_{\mathrm{m}}(\zeta)$~[\si{\becquerel \per \kilo \gram}] is the activity per unit soil mass at mass depth $\zeta$~[\si{\gram \per \square \centi \metre}], $A_{\mathrm{m},0}$~[\si{\becquerel \per \kilo \gram}] is the activity per unit mass at the ground surface, and $\beta$~[\si{\gram \per \square \centi \metre}] is the relaxation mass per unit area that quantifies the depth of fallout penetration into the ground. The mass depth is
\begin{equation}
\zeta=\int_{0}^{z}\rho_\mathrm{s}(z')\,\mathrm{d}z'\,,
\end{equation}
\noindent where $z$~[\si{\centi \metre}] is the depth below the ground surface.

\begin{figure}
 \includegraphics[width=0.4\textwidth]{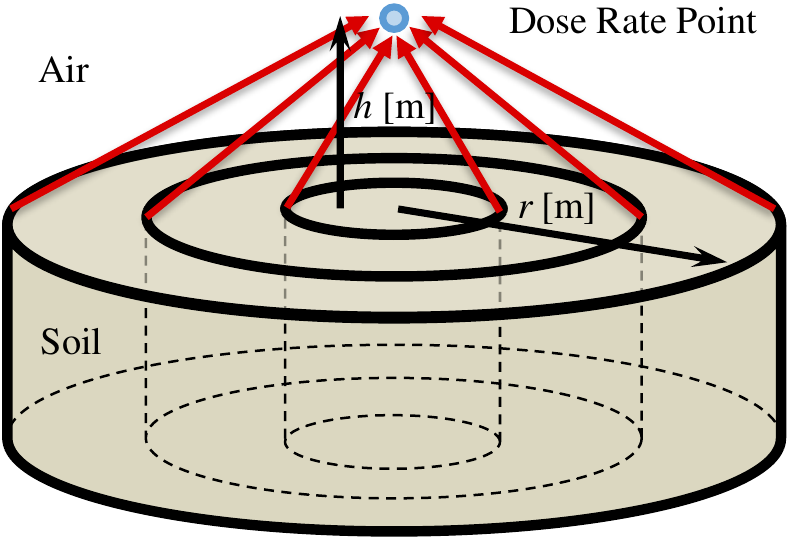}
 \caption{\label{fig:geom}The field of view for $\dot{H}^*(10)$ was calculated in terms of the contribution from the decay of radionuclides within increasing radii volumes of ground beneath the dose rate point.}
\end{figure}

For computational efficiency, the radiation source volume was scaled to a line within the soil at the centre of the simulation space, and the Monte Carlo tally regions to circular planes at fixed height above the ground and with varying radii~\cite{Namito2012}. This set-up enabled fast calculation of a field of view in terms of contributions from increasing radii cylindrical volumes of soil centred beneath the air dose rate location. Fig.~\ref{fig:geom} shows the real geometry achieved by this simulation method. The tally response function calculated was the ambient dose equivalent rate, $\dot{H}^*(10)$~[\si{\micro\sievert\per\hour}]~\cite{ICRP74}.

Two limitations of simulating the infinite half-space geometry are as follows. First, as the model assumes perfectly flat land it does not account for the radiation shielding provided by any topographic features on the land, such as slopes and mountains, or shielding by vegetation and buildings. The effect of shielding is to decrease the field of view, so the results presented here can be considered as upper bounds for field of view in settings where shielding features are present.

The second limitation is that the model assumes a perfectly uniform distribution of the radionuclides across the land surface. This approximation may be overly simplistic if geological or soil conditions lead to different concentrations of natural radionuclides across an area. Or for anthropogenic radionuclides, atmospheric fallout and subsequent radionuclide absorption and redistribution may be quite heterogeneous, again depending on local conditions. These source distribution effects can widen or narrow the air dose rate field of view depending on a case-by-case basis. The reader should be aware that heterogeneous source distributions may be important in setting the field of view at some locations.

\section{\label{sec:Results}Results}

\begin{figure}
 \includegraphics[width=0.45\textwidth]{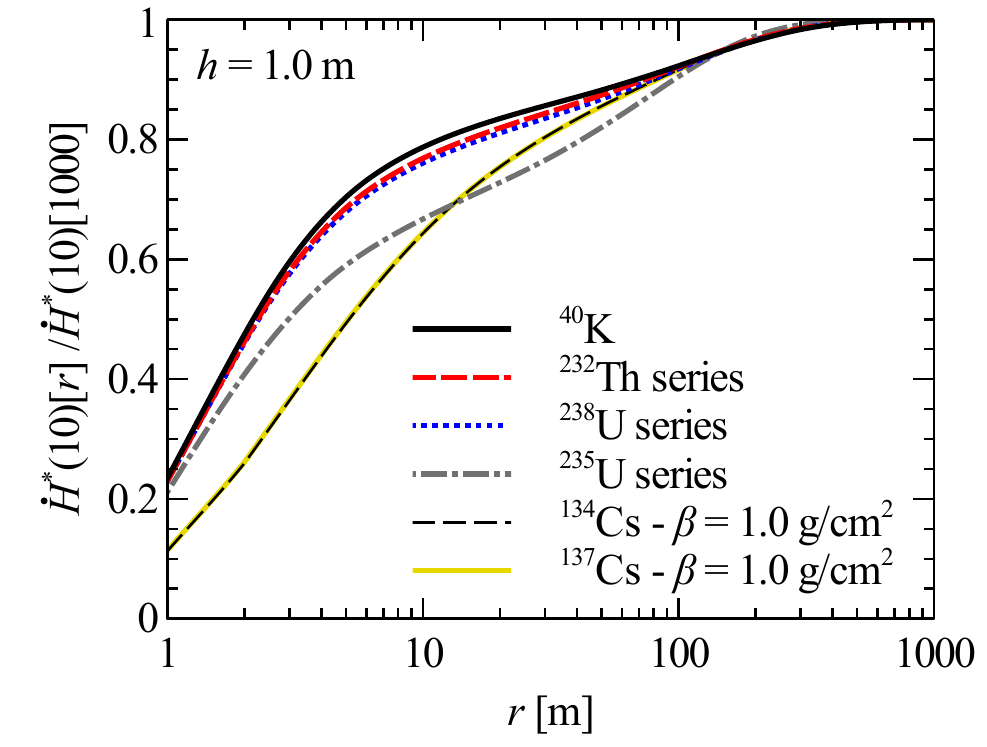}
 \caption{\label{fig:fov_natural}Comparing the field of view from natural radionuclides, \textsuperscript{134}Cs and \textsuperscript{137}Cs.}
\end{figure}
The field of view from natural radionuclides distributed homogeneously within the ground and for \textsuperscript{134}Cs and \textsuperscript{137}Cs exponentially distributed with depth ($\beta=$~\SI{1.0}{\gram \per \square \centi \metre}) are shown in Fig.~\ref{fig:fov_natural}. The graph shows the fraction of $\dot{H}^*(10)$ at \SI{1}{\metre} above the ground attributable to each of the sources within a radius $r$~[\si{\metre}] cylindrical volume of soil below the dose rate point. The data are plotted relative to the ambient dose equivalent rate from a \SI{1000}{\metre} radius volume, which is a sufficiently large range for $\dot{H}^*(10)$ to reach its asymptotic limit to three significant figures. 

The length scale of the field of view is of comparable order of magnitude for all the nuclides. Around \SIrange{60}{80}{\percent} of the radiation contributing to $\dot{H}^*(10)$ originates from within \SI{10}{\metre} around the dose rate point. Slowly increasing tails on the graphs mean that it takes \SI{100}{\metre} for $\dot{H}^*(10)$ to reach \SI{90}{\percent} of its infinite limit. Radiation contributing to the final \SI{10}{\percent} of the limiting $\dot{H}^*(10)$ originates more than \SI{100}{\metre} from the dose rate location.

To compare the fields of view of the natural radionuclides with radiocaesium, we consider the source radius contributing to \SI{60}{\percent} of $\dot{H}^*(10)$. The field of view is narrower for \textsuperscript{40}K, the \textsuperscript{232}Th series and the \textsuperscript{238}U series ($r \simeq 4$~\si{\metre}) than for radiocaesium ($r = 8$~\si{\metre}). This is a consequence of the different distributions of the radionuclides within the ground, and because higher energy decay photons (Fig.~\ref{fig:emission_spectra}) make the dominant contribution to $\dot{H}^*(10)$ for \textsuperscript{40}K, the \textsuperscript{232}Th series and the \textsuperscript{238}U series than for \textsuperscript{134}Cs and \textsuperscript{137}Cs.

The field of view for \SI{60}{\percent} of $\dot{H}^*(10)$ is wider for the \textsuperscript{235}U series ($r = 5.5$~\si{\metre}) than for the other natural nuclides. This is because of the lower energy decay photons emitted by the \textsuperscript{235}U series (typically \SI{<500}{\kilo \electronvolt} - Fig.~\ref{fig:emission_spectra}). It should be noted that in the environment, the \textsuperscript{235}U series makes a minimal contribution to the total ambient dose equivalent rate from natural terrestrial sources~\cite{UNSCEAR2000}. This is because of its low abundance in natural uranium ($\simeq 0.7$~\si{\percent} by mass).

The field of view for the air dose rate from natural terrestrial sources is unlikely to change significantly depending on the relative concentrations of \textsuperscript{40}K, the \textsuperscript{232}Th series and the \textsuperscript{238}U series radioisotopes in the ground. This is because the fields of view for each of these three components are similar (black, red and blue lines in Fig.~\ref{fig:fov_natural}).

\begin{figure}
 \includegraphics[width=0.45\textwidth]{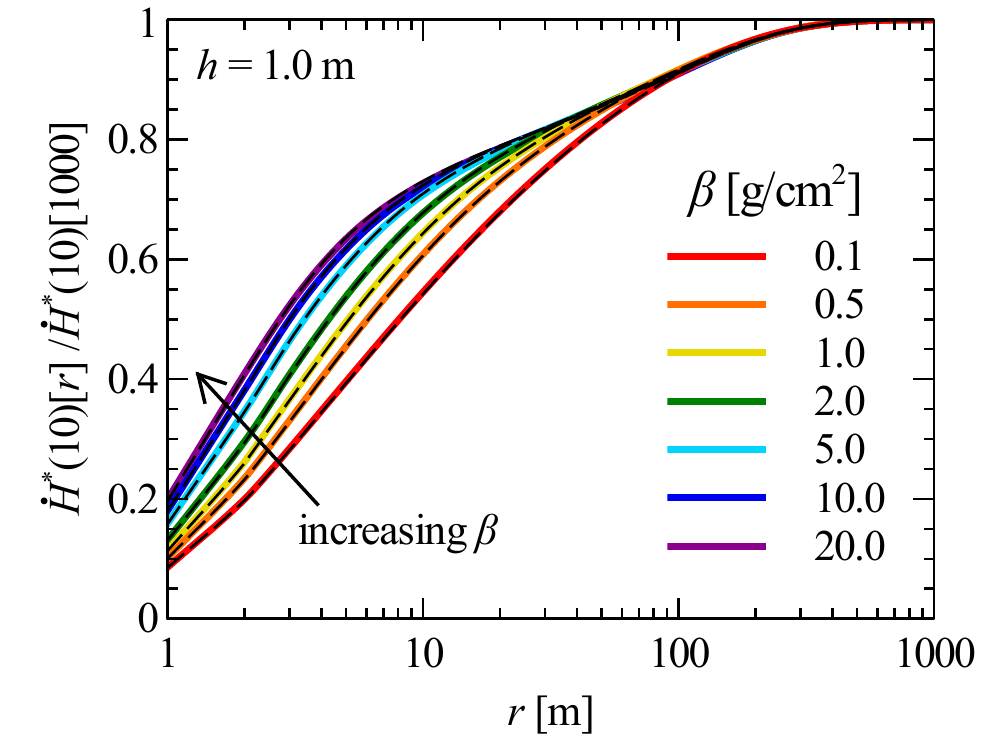}
 \caption{\label{fig:fov_Cs_beta}The field of view from \textsuperscript{134}Cs and \textsuperscript{137}Cs with varying $\beta$.}
\end{figure}

Fig.~\ref{fig:fov_Cs_beta} shows the effect of fallout penetration into the soil on the field of view from \textsuperscript{134}Cs and \textsuperscript{137}Cs. The relaxation mass per unit area is shown for the range \SIrange{0.1}{20.0}{\gram \per \square \centi \metre}. $\beta$ is typically between \SIrange{0.1}{3.0}{\gram \per \square \centi \metre} in the first year after fallout deposition, increasing to \SIrange{1.0}{7.0}{\gram \per \square \centi \metre} after several years, and then in the range \SIrange{2.0}{20.0}{\gram \per \square \centi \metre} for fallout deposited greater than 10 years previously~\cite{ICRU53}. When considering the source radius contributing to \SI{60}{\percent} of $\dot{H}^*(10)$, the field of view narrows as fallout radiocaesium weathers deeper into the ground ($r$ decreases from \SI{13.5}{\metre} to \SI{4.5}{\metre} when $\beta$ increases between \SIrange{0.1}{20.0}{\gram \per \square \centi \metre}, Fig.~\ref{fig:fov_Cs_beta}).

\begin{figure}
 \includegraphics[width=0.45\textwidth]{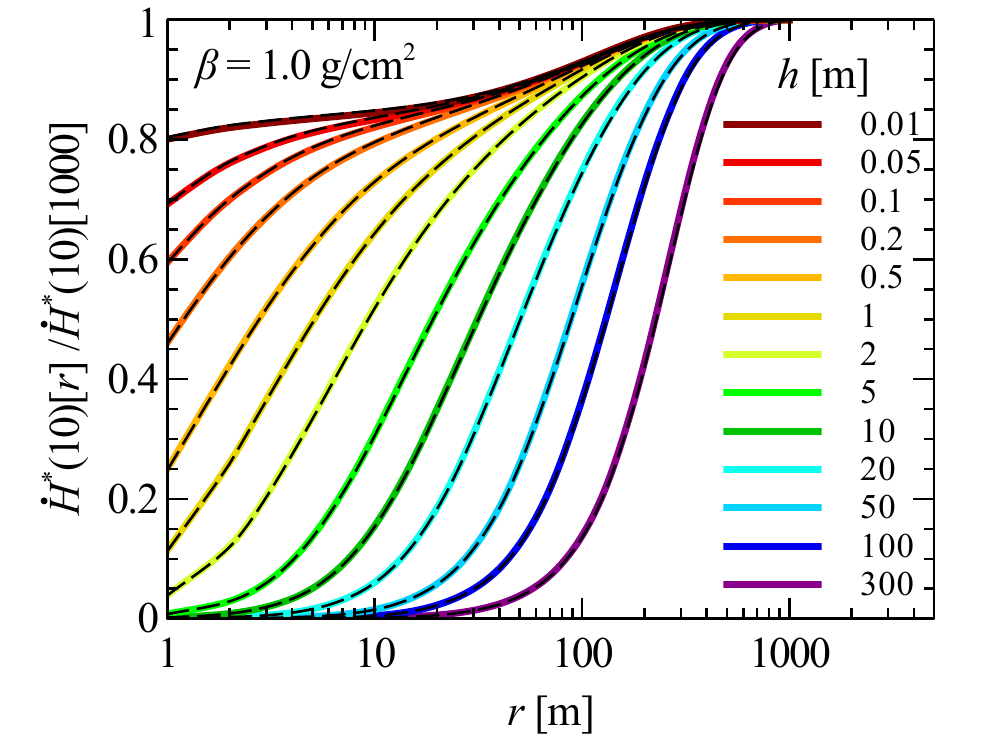}
 \caption{\label{fig:fov_Cs_h}The field of view from \textsuperscript{134}Cs and \textsuperscript{137}Cs at various heights above the ground.}
\end{figure}

The effect of elevation above the ground on field of view for $\dot{H}^*(10)$ from \textsuperscript{134}Cs and \textsuperscript{137}Cs is shown in Fig.~\ref{fig:fov_Cs_h}. Elevating the dose rate point changes the relative distance between different areas of the ground and the dose rate point, and leads to more shielding of the radiation by air. The field of view always widens with increasing height of the dose rate point. It is narrow when the dose rate point is located close to the ground (less than \SI{1}{\metre} for \SI{60}{\percent} of $\dot{H}^*(10)$ when $h=0.01$~\si{\metre}). It extends to over \SI{100}{\metre} for the highest elevations studied ($h \geq 100$~\si{\metre}). 

In all cases the tail of the distributions mean that \textsuperscript{134}Cs and \textsuperscript{137}Cs at large distances makes a significant contribution to the dose rate. Fallout more than \SI{100}{\metre} from the dose rate point needs to be considered for $\dot{H}^*(10)$ to reach its asymptotic limit at all dose rate elevations above the ground.

Figs.~\ref{fig:fov_natural}--\ref{fig:fov_Cs_h} show that the fields of view from \textsuperscript{134}Cs and \textsuperscript{137}Cs are almost identical. This is because the dominant photons contributing to $\dot{H}^*(10)$ emitted by both isotopes have similar energies (\SIrange{600}{800}{\kilo \electronvolt} -- Fig.~\ref{fig:emission_spectra}).

\section{\label{sec:Applications}Example Applications}

\subsection{\label{ssec:KURAMA}Interpreting KURAMA Car and Bus Survey Data}

KURAMA and KURAMA-II are vehicle-borne radiation survey systems deployed on cars and buses within Fukushima Prefecture and across Japan to measure air dose rates~\cite{Tanigaki2013,Andoh2014,Tsuda2014,Tanigaki2015}. They consist of a NaI/CsI scintillator to detect gamma rays, a GPS unit to determine measurement locations, a control computer, and a 3G network connection to transmit results.

\begin{figure}
 \includegraphics[width=0.45\textwidth]{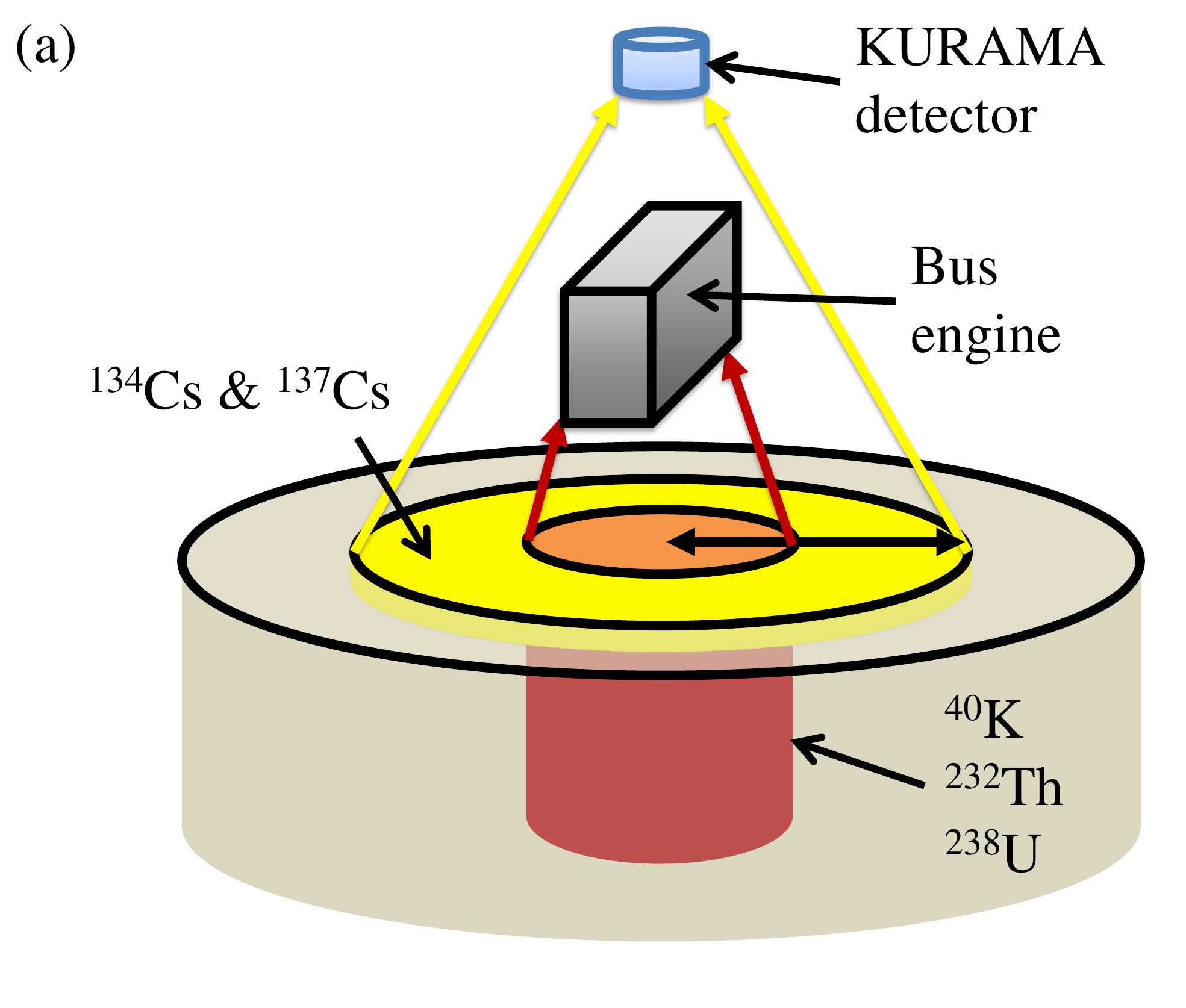} \includegraphics[width=0.45\textwidth]{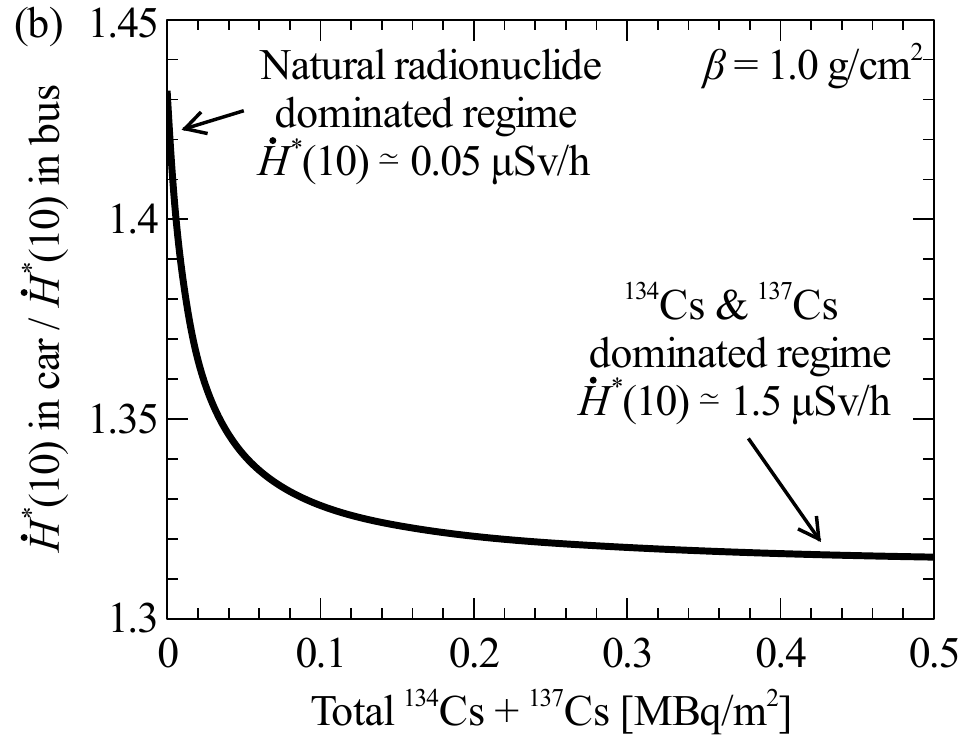}
 \caption{\label{fig:bus_car_ratio}(a) The bus engine shields a greater portion of the field of view of natural radionuclides than for \textsuperscript{134}Cs and \textsuperscript{137}Cs. Shaded red and yellow volumes illustrate natural and \textsuperscript{134/137}Cs radionuclides in respective fields of view. (b) Ratio of $\dot{H}^*(10)$ at the KURAMA position within a car to within a bus, as a function of environmental radiocaesium levels. Inset text on graph lists outdoor $\dot{H}^*(10)$ values. \textsuperscript{134}Cs/\textsuperscript{137}Cs activity ratio is \num{0.57}, applicable on January 01, 2013.}
\end{figure}

In cars, the devices are mounted either above the rear right-side door or behind the headrest of the rear seat~\cite{Tanigaki2013,Andoh2014}. In public buses, the systems are placed above the engine bay at the rear of the bus~\cite{Tanigaki2015}. These geometries mean that the systems respond differently to radiation from \textsuperscript{134}Cs and \textsuperscript{137}Cs fallout, and to radiation from natural radionuclides when operated in different types of vehicle.

The difference is due to differences in the fields of view between the anthropogenic and the natural radionuclides, and the fact that the bus engine acts as a shield. The effect is depicted in Fig.~\ref{fig:bus_car_ratio}(a). The narrower field of view from natural radionuclides than from \textsuperscript{134}Cs and \textsuperscript{137}Cs means that a greater portion of the field of view from natural radionuclides is shielded by the bus engine.

The effect bears out as a different response of the KURAMA systems when mounted in cars and buses between low dose rate areas, where radiation from natural terrestrial radionuclides dominates (\SI{\approx 0.05}{\micro \sievert \per \hour}), and high dose rate areas (\SI{>0.25}{\micro \sievert \per \hour}) with large amounts of \textsuperscript{134}Cs and \textsuperscript{137}Cs fallout. The graph in Fig.~\ref{fig:bus_car_ratio}(b) shows how the ratio of the air dose rate within a car to the dose rate above a bus engine bay depends on the level of \textsuperscript{134}Cs and \textsuperscript{137}Cs fallout in the environment. The data in Fig.~\ref{fig:bus_car_ratio}(b) were generated by PHITS simulations of car and bus geometries placed within the half-space geometry.

In the low dose rate regime, the ratio is high as the dose rate in the bus is lowered by the shielding of gamma rays from natural radionuclides by the bus engine. However, at high dose rates radiocaesium is dominant, and its wider field of view means that the shielding by the bus engine is less important. Thus, the ratio of the air dose rate in the car and the bus is smaller. This effect is also seen in real data from KURAMA operation on vehicles in Fukushima.

The ratio of the dose rates in the car and bus becomes less sensitive to the fallout levels as caesium migrates deeper into the ground. This is because the field of view of radiocaesium narrows with deeper migration (Fig.~\ref{fig:fov_Cs_beta}) and becomes more comparable to the field of view from natural radionuclides. The shielding effect of the bus engine is therefore less important.

\subsection{\label{ssec:remediation}The Effectiveness of Land Remediation}

\begin{figure}
 \includegraphics[width=0.45\textwidth]{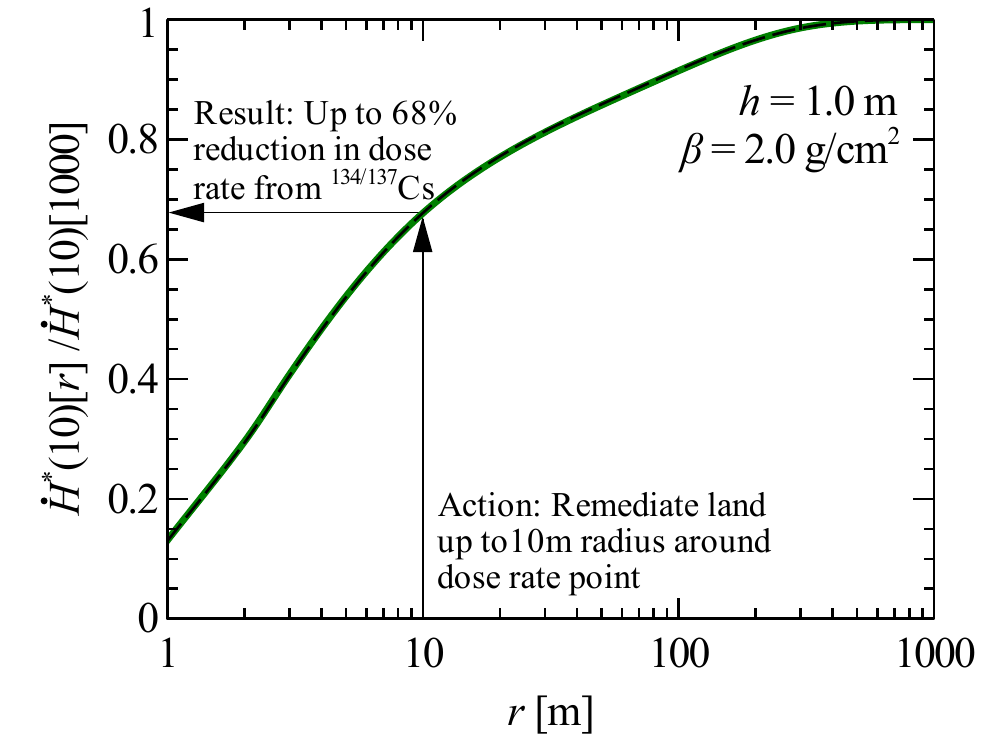}
 \caption{\label{fig:fov_Cs_remediation}Using the field of view to evaluate the effectiveness of remediation for reducing $\dot{H}^*(10)$.}
\end{figure}

Field of view graphs like Figs.~\ref{fig:fov_natural}--\ref{fig:fov_Cs_h} can be used to quickly evaluate the effectiveness of land remediation for reducing air dose rates. Fig.~\ref{fig:fov_Cs_remediation} demonstrates the technique.

The field of view results show that \SI{68}{\percent} of the ambient dose equivalent rate attributable to the radiocaesium fallout comes from \textsuperscript{134}Cs and \textsuperscript{137}Cs within a \SI{10}{\metre} radius area on the ground below the dose rate point. If the land within this area were remediated perfectly, so that no \textsuperscript{134}Cs and \textsuperscript{137}Cs remained afterwards, the component of $\dot{H}^*(10)$ attributable to radiocaesium would decrease by \SI{68}{\percent}. 

It is also possible to account for the effectiveness of the remediation through the decontamination factor (DF). DF is defined as the ratio of the caesium radioactivity in the land before and after remediation. DF values vary between different remediation methods. For $\text{DF}=20$, \SI{5}{\percent} of the radioactivity will remain in the land after remediation, and hence the reduction in the \textsuperscript{134}Cs and \textsuperscript{137}Cs component of $\dot{H}^*(10)$ would be $68\,\% \cdot (100\,\% - 5\,\%) = 64.6\,\%$.

The field of view graphs thus allow quick calculations for the order of magnitude of reduction in a dose rate that can be achieved by remediation. More accurate calculations for the effectiveness of remediation can be obtained using a specialist tool to model the land type, topography and local area of the remediation target. An example of such a tool is the calculation system for the estimation of decontamination effects (CDE)~\cite{Satoh2014}.

\section{\label{sec:Conclusions}Conclusions}

PHITS simulations were used to characterize the field of view from natural terrestrial radionuclides and anthropogenic caesium fallout in the environment. \textsuperscript{134}Cs and \textsuperscript{137}Cs fallout has a wider field of view than the natural radionuclides (excluding \textsuperscript{235}U series) due to different source positions and characteristic gamma ray energies. The field of view from \textsuperscript{134}Cs and \textsuperscript{137}Cs fallout narrows as the radiocaesium migrates deeper into the ground. The \textsuperscript{134}Cs and \textsuperscript{137}Cs field of view widens with height of the air dose rate location above the ground.

The differences in the field of view between natural radionuclides and \textsuperscript{134}Cs and \textsuperscript{137}Cs can explain different KURAMA detector responses between low dose rate and high dose rate areas when operated above a bus engine bay. Field of view graphs can be used to make quick calculations for the reduction in air dose rates before and after land remediation.

\begin{acknowledgments}

We acknowledge fruitful discussions with M.~Tanigaki, D.~C.~W.~Sanderson and JAEA colleagues. Calculations were performed on JAEA's BX900 supercomputer.

\end{acknowledgments}

\bibliography{Revitalizing_Fukushima_15_Proceedings}

\begin{thebibliography}{20}%
\makeatletter
\providecommand \@ifxundefined [1]{%
 \@ifx{#1\undefined}
}%
\providecommand \@ifnum [1]{%
 \ifnum #1\expandafter \@firstoftwo
 \else \expandafter \@secondoftwo
 \fi
}%
\providecommand \@ifx [1]{%
 \ifx #1\expandafter \@firstoftwo
 \else \expandafter \@secondoftwo
 \fi
}%
\providecommand \natexlab [1]{#1}%
\providecommand \enquote  [1]{``#1''}%
\providecommand \bibnamefont  [1]{#1}%
\providecommand \bibfnamefont [1]{#1}%
\providecommand \citenamefont [1]{#1}%
\providecommand \href@noop [0]{\@secondoftwo}%
\providecommand \href [0]{\begingroup \@sanitize@url \@href}%
\providecommand \@href[1]{\@@startlink{#1}\@@href}%
\providecommand \@@href[1]{\endgroup#1\@@endlink}%
\providecommand \@sanitize@url [0]{\catcode `\\12\catcode `\$12\catcode
  `\&12\catcode `\#12\catcode `\^12\catcode `\_12\catcode `\%12\relax}%
\providecommand \@@startlink[1]{}%
\providecommand \@@endlink[0]{}%
\providecommand \url  [0]{\begingroup\@sanitize@url \@url }%
\providecommand \@url [1]{\endgroup\@href {#1}{\urlprefix }}%
\providecommand \urlprefix  [0]{URL }%
\providecommand \Eprint [0]{\href }%
\providecommand \doibase [0]{http://dx.doi.org/}%
\providecommand \selectlanguage [0]{\@gobble}%
\providecommand \bibinfo  [0]{\@secondoftwo}%
\providecommand \bibfield  [0]{\@secondoftwo}%
\providecommand \translation [1]{[#1]}%
\providecommand \BibitemOpen [0]{}%
\providecommand \bibitemStop [0]{}%
\providecommand \bibitemNoStop [0]{.\EOS\space}%
\providecommand \EOS [0]{\spacefactor3000\relax}%
\providecommand \BibitemShut  [1]{\csname bibitem#1\endcsname}%
\let\auto@bib@innerbib\@empty
\bibitem [{\citenamefont {{ICRP Publ. 74}}(1996)}]{ICRP74}%
  \BibitemOpen
  \bibfield  {author} {\bibinfo {author} {\bibnamefont {{ICRP Publ. 74}}},\
  }\href {\doibase 10.1016/S0146-6453(96)90001-9} {\bibfield  {journal}
  {\bibinfo  {journal} {Ann. ICRP}\ }\textbf {\bibinfo {volume} {26}},\
  \bibinfo {pages} {1} (\bibinfo {year} {1996})}\BibitemShut {NoStop}%
\bibitem [{\citenamefont {Hubbell}\ and\ \citenamefont
  {Seltzer}(2004)}]{Hubbell2004}%
  \BibitemOpen
  \bibfield  {author} {\bibinfo {author} {\bibfnamefont {J.~H.}\ \bibnamefont
  {Hubbell}}\ and\ \bibinfo {author} {\bibfnamefont {S.~M.}\ \bibnamefont
  {Seltzer}},\ }\href {http://www.nist.gov/pml/data/xraycoef/} {\enquote
  {\bibinfo {title} {{Tables of X-Ray Mass Attenuation Coefficients and Mass
  Energy-Absorption Coefficients from 1 keV to 20 MeV for Elements Z = 1 to 92
  and 48 Additional Substances of Dosimetric Interest}},}\ } (\bibinfo {year}
  {2004})\BibitemShut {NoStop}%
\bibitem [{\citenamefont {Allyson}(1994)}]{Allyson1994}%
  \BibitemOpen
  \bibfield  {author} {\bibinfo {author} {\bibfnamefont {J.~D.}\ \bibnamefont
  {Allyson}},\ }\emph {\bibinfo {title} {{Environmental {$\gamma$}-ray
  Spectrometry: Simulation of Absolute Calibration of In-Situ and Airborne
  Spectrometers for Natural and Anthropogenic Sources}}},\ \href
  {http://theses.gla.ac.uk/2028/1/1994allysonphd.pdf} {Ph.D. thesis},\ \bibinfo
   {school} {The University of Glasgow} (\bibinfo {year} {1994})\BibitemShut
  {NoStop}%
\bibitem [{\citenamefont {Tyler}\ \emph {et~al.}(1996)\citenamefont {Tyler},
  \citenamefont {Sanderson}, \citenamefont {Scott},\ and\ \citenamefont
  {Allyson}}]{Tyler1996}%
  \BibitemOpen
  \bibfield  {author} {\bibinfo {author} {\bibfnamefont {A.~N.}\ \bibnamefont
  {Tyler}}, \bibinfo {author} {\bibfnamefont {D.~C.~W.}\ \bibnamefont
  {Sanderson}}, \bibinfo {author} {\bibfnamefont {E.~M.}\ \bibnamefont
  {Scott}}, \ and\ \bibinfo {author} {\bibfnamefont {J.~D.}\ \bibnamefont
  {Allyson}},\ }\href {\doibase 10.1016/0265-931X(95)00097-T} {\bibfield
  {journal} {\bibinfo  {journal} {J. Environ. Radioactiv.}\ }\textbf {\bibinfo
  {volume} {33}},\ \bibinfo {pages} {213} (\bibinfo {year} {1996})}\BibitemShut
  {NoStop}%
\bibitem [{\citenamefont {Iwamoto}\ \emph {et~al.}(2011)\citenamefont
  {Iwamoto}, \citenamefont {Satoh}, \citenamefont {Endo}, \citenamefont
  {Sakamoto}, \citenamefont {Kureta},\ and\ \citenamefont
  {Kugo}}]{Iwamoto2011}%
  \BibitemOpen
  \bibfield  {author} {\bibinfo {author} {\bibfnamefont {Y.}~\bibnamefont
  {Iwamoto}}, \bibinfo {author} {\bibfnamefont {D.}~\bibnamefont {Satoh}},
  \bibinfo {author} {\bibfnamefont {A.}~\bibnamefont {Endo}}, \bibinfo {author}
  {\bibfnamefont {Y.}~\bibnamefont {Sakamoto}}, \bibinfo {author}
  {\bibfnamefont {M.}~\bibnamefont {Kureta}}, \ and\ \bibinfo {author}
  {\bibfnamefont {T.}~\bibnamefont {Kugo}},\ }\href {\doibase
  10.11484/jaea-technology-2011-026} {\emph {\bibinfo {title} {{Study on Soil
  Decontamination and Dose Rate Reduction Effect}}}},\ \bibinfo {type} {Tech.
  Rep.}\ \bibinfo {number} {JAEA-Technology 2011-026}\ (\bibinfo  {institution}
  {Japan Atomic Energy Agency},\ \bibinfo {year} {2011})\BibitemShut {NoStop}%
\bibitem [{\citenamefont {Saito}(1991)}]{Saito1991}%
  \BibitemOpen
  \bibfield  {author} {\bibinfo {author} {\bibfnamefont {K.}~\bibnamefont
  {Saito}},\ }\href {http://rpd.oxfordjournals.org/content/35/1/31.short}
  {\bibfield  {journal} {\bibinfo  {journal} {Radiat. Prot. Dosim.}\ }\textbf
  {\bibinfo {volume} {35}},\ \bibinfo {pages} {31} (\bibinfo {year}
  {1991})}\BibitemShut {NoStop}%
\bibitem [{\citenamefont {Saito}\ and\ \citenamefont
  {Jacob}(1995)}]{Saito1995}%
  \BibitemOpen
  \bibfield  {author} {\bibinfo {author} {\bibfnamefont {K.}~\bibnamefont
  {Saito}}\ and\ \bibinfo {author} {\bibfnamefont {P.}~\bibnamefont {Jacob}},\
  }\href {http://rpd.oxfordjournals.org/content/58/1/29.abstract} {\bibfield
  {journal} {\bibinfo  {journal} {Radiat. Prot. Dosim.}\ }\textbf {\bibinfo
  {volume} {58}},\ \bibinfo {pages} {29} (\bibinfo {year} {1995})}\BibitemShut
  {NoStop}%
\bibitem [{\citenamefont {Saito}\ and\ \citenamefont
  {Jacob}(1998)}]{Saito1998}%
  \BibitemOpen
  \bibfield  {author} {\bibinfo {author} {\bibfnamefont {K.}~\bibnamefont
  {Saito}}\ and\ \bibinfo {author} {\bibfnamefont {P.}~\bibnamefont {Jacob}},\
  }\href
  {http://jolissrch-inter.tokai-sc.jaea.go.jp/pdfdata/JAERI-Data-Code-98-001.pdf}
  {\emph {\bibinfo {title} {{Fundamental Data on Environmental Gamma-Ray Fields
  in the Air due to Sources in the Ground}}}},\ \bibinfo {type} {Tech. Rep.}\
  \bibinfo {number} {JAERI-Data/Code 98-001}\ (\bibinfo {year}
  {1998})\BibitemShut {NoStop}%
\bibitem [{\citenamefont {Sato}\ \emph {et~al.}(2013)\citenamefont {Sato},
  \citenamefont {Niita}, \citenamefont {Matsuda}, \citenamefont {Hashimoto},
  \citenamefont {Iwamoto}, \citenamefont {Noda}, \citenamefont {Ogawa},
  \citenamefont {Iwase}, \citenamefont {Nakashima}, \citenamefont {Fukahori},
  \citenamefont {Okumura}, \citenamefont {Kai}, \citenamefont {Chiba},
  \citenamefont {Furuta},\ and\ \citenamefont {Sihver}}]{Sato2013}%
  \BibitemOpen
  \bibfield  {author} {\bibinfo {author} {\bibfnamefont {T.}~\bibnamefont
  {Sato}}, \bibinfo {author} {\bibfnamefont {K.}~\bibnamefont {Niita}},
  \bibinfo {author} {\bibfnamefont {N.}~\bibnamefont {Matsuda}}, \bibinfo
  {author} {\bibfnamefont {S.}~\bibnamefont {Hashimoto}}, \bibinfo {author}
  {\bibfnamefont {Y.}~\bibnamefont {Iwamoto}}, \bibinfo {author} {\bibfnamefont
  {S.}~\bibnamefont {Noda}}, \bibinfo {author} {\bibfnamefont {T.}~\bibnamefont
  {Ogawa}}, \bibinfo {author} {\bibfnamefont {H.}~\bibnamefont {Iwase}},
  \bibinfo {author} {\bibfnamefont {H.}~\bibnamefont {Nakashima}}, \bibinfo
  {author} {\bibfnamefont {T.}~\bibnamefont {Fukahori}}, \bibinfo {author}
  {\bibfnamefont {K.}~\bibnamefont {Okumura}}, \bibinfo {author} {\bibfnamefont
  {T.}~\bibnamefont {Kai}}, \bibinfo {author} {\bibfnamefont {S.}~\bibnamefont
  {Chiba}}, \bibinfo {author} {\bibfnamefont {T.}~\bibnamefont {Furuta}}, \
  and\ \bibinfo {author} {\bibfnamefont {L.}~\bibnamefont {Sihver}},\ }\href
  {\doibase 10.1080/00223131.2013.814553} {\bibfield  {journal} {\bibinfo
  {journal} {J. Nucl. Sci. Technol.}\ }\textbf {\bibinfo {volume} {50}},\
  \bibinfo {pages} {913} (\bibinfo {year} {2013})}\BibitemShut {NoStop}%
\bibitem [{\citenamefont {{ICRU Publ. 53}}(1994)}]{ICRU53}%
  \BibitemOpen
  \bibfield  {author} {\bibinfo {author} {\bibnamefont {{ICRU Publ. 53}}},\
  }\href
  {http://www.icru.org/home/reports/gamma-ray-spectrometry-in-the-environment-report-53}
  {\emph {\bibinfo {title} {{Gamma-Ray Spectrometry in the Environment}}}}\
  (\bibinfo {address} {Bethesda},\ \bibinfo {year} {1994})\BibitemShut
  {NoStop}%
\bibitem [{\citenamefont {Eckerman}\ and\ \citenamefont
  {Ryman}(1993)}]{Eckerman1993}%
  \BibitemOpen
  \bibfield  {author} {\bibinfo {author} {\bibfnamefont {K.~F.}\ \bibnamefont
  {Eckerman}}\ and\ \bibinfo {author} {\bibfnamefont {J.~C.}\ \bibnamefont
  {Ryman}},\ }\href
  {http://www.epa.gov/radiation/docs/federal/402-r-93-081.pdf} {\emph {\bibinfo
  {title} {{External Exposure to Radionuclides in Air, Water, and Soil}}}},\
  \bibinfo {type} {Tech. Rep.}\ \bibinfo {number} {Federal Guidance Report No.
  12, EPA-402-R-93-081}\ (\bibinfo  {institution} {U.S. Environmental
  Protection Agency},\ \bibinfo {year} {1993})\BibitemShut {NoStop}%
\bibitem [{\citenamefont {{ICRP Publ. 107}}(2008)}]{ICRP107}%
  \BibitemOpen
  \bibfield  {author} {\bibinfo {author} {\bibnamefont {{ICRP Publ. 107}}},\
  }\href {\doibase 10.1016/j.icrp.2008.10.003} {\bibfield  {journal} {\bibinfo
  {journal} {Ann. ICRP}\ }\textbf {\bibinfo {volume} {38}},\ \bibinfo {pages}
  {119} (\bibinfo {year} {2008})}\BibitemShut {NoStop}%
\bibitem [{\citenamefont {NuDat2}(2014)}]{NuDat2}%
  \BibitemOpen
  \bibfield  {author} {\bibinfo {author} {\bibnamefont {NuDat2}},\ }\href
  {http://www.nndc.bnl.gov/nudat2/} {\enquote {\bibinfo {title} {{Software to
  search and plot nuclear structure and decay data interactively. Employs data
  from the Evaluated Nuclear Structure Data File (ENSDF).}}}\ } (\bibinfo
  {year} {2014})\BibitemShut {NoStop}%
\bibitem [{\citenamefont {Namito}\ \emph {et~al.}(2012)\citenamefont {Namito},
  \citenamefont {Nakamura}, \citenamefont {Toyoda}, \citenamefont {Iijima},
  \citenamefont {Iwase}, \citenamefont {Ban},\ and\ \citenamefont
  {Hirayama}}]{Namito2012}%
  \BibitemOpen
  \bibfield  {author} {\bibinfo {author} {\bibfnamefont {Y.}~\bibnamefont
  {Namito}}, \bibinfo {author} {\bibfnamefont {H.}~\bibnamefont {Nakamura}},
  \bibinfo {author} {\bibfnamefont {A.}~\bibnamefont {Toyoda}}, \bibinfo
  {author} {\bibfnamefont {K.}~\bibnamefont {Iijima}}, \bibinfo {author}
  {\bibfnamefont {H.}~\bibnamefont {Iwase}}, \bibinfo {author} {\bibfnamefont
  {S.}~\bibnamefont {Ban}}, \ and\ \bibinfo {author} {\bibfnamefont
  {H.}~\bibnamefont {Hirayama}},\ }\href {\doibase
  10.1080/00223131.2011.649079} {\bibfield  {journal} {\bibinfo  {journal} {J.
  Nucl. Sci. Technol.}\ }\textbf {\bibinfo {volume} {49}},\ \bibinfo {pages}
  {167} (\bibinfo {year} {2012})}\BibitemShut {NoStop}%
\bibitem [{\citenamefont {UNSCEAR}(2000)}]{UNSCEAR2000}%
  \BibitemOpen
  \bibfield  {author} {\bibinfo {author} {\bibnamefont {UNSCEAR}},\ }\href
  {http://www.unscear.org/unscear/en/publications/2000_1.html} {\emph {\bibinfo
  {title} {{Sources, Effects and Risks of Ionizing Radiation, UNSCEAR 2000
  Report, Volume I: Sources}}}}\ (\bibinfo {year} {2000})\BibitemShut {NoStop}%
\bibitem [{\citenamefont {Tanigaki}\ \emph {et~al.}(2013)\citenamefont
  {Tanigaki}, \citenamefont {Okumura}, \citenamefont {Takamiya}, \citenamefont
  {Sato}, \citenamefont {Yoshino},\ and\ \citenamefont
  {Yamana}}]{Tanigaki2013}%
  \BibitemOpen
  \bibfield  {author} {\bibinfo {author} {\bibfnamefont {M.}~\bibnamefont
  {Tanigaki}}, \bibinfo {author} {\bibfnamefont {R.}~\bibnamefont {Okumura}},
  \bibinfo {author} {\bibfnamefont {K.}~\bibnamefont {Takamiya}}, \bibinfo
  {author} {\bibfnamefont {N.}~\bibnamefont {Sato}}, \bibinfo {author}
  {\bibfnamefont {H.}~\bibnamefont {Yoshino}}, \ and\ \bibinfo {author}
  {\bibfnamefont {H.}~\bibnamefont {Yamana}},\ }\href {\doibase
  10.1016/j.nima.2013.05.059} {\bibfield  {journal} {\bibinfo  {journal} {Nucl.
  Instrum. Methods Phys. Res. Sect. A}\ }\textbf {\bibinfo {volume} {726}},\
  \bibinfo {pages} {162} (\bibinfo {year} {2013})}\BibitemShut {NoStop}%
\bibitem [{\citenamefont {Andoh}\ \emph {et~al.}(2014)\citenamefont {Andoh},
  \citenamefont {Nakahara}, \citenamefont {Tsuda}, \citenamefont {Yoshida},
  \citenamefont {Matsuda}, \citenamefont {Takahashi}, \citenamefont {Mikami},
  \citenamefont {Kinouchi}, \citenamefont {Sato}, \citenamefont {Tanigaki},
  \citenamefont {Takamiya}, \citenamefont {Sato}, \citenamefont {Okumura},
  \citenamefont {Uchihori},\ and\ \citenamefont {Saito}}]{Andoh2014}%
  \BibitemOpen
  \bibfield  {author} {\bibinfo {author} {\bibfnamefont {M.}~\bibnamefont
  {Andoh}}, \bibinfo {author} {\bibfnamefont {Y.}~\bibnamefont {Nakahara}},
  \bibinfo {author} {\bibfnamefont {S.}~\bibnamefont {Tsuda}}, \bibinfo
  {author} {\bibfnamefont {T.}~\bibnamefont {Yoshida}}, \bibinfo {author}
  {\bibfnamefont {N.}~\bibnamefont {Matsuda}}, \bibinfo {author} {\bibfnamefont
  {F.}~\bibnamefont {Takahashi}}, \bibinfo {author} {\bibfnamefont
  {S.}~\bibnamefont {Mikami}}, \bibinfo {author} {\bibfnamefont
  {N.}~\bibnamefont {Kinouchi}}, \bibinfo {author} {\bibfnamefont
  {T.}~\bibnamefont {Sato}}, \bibinfo {author} {\bibfnamefont {M.}~\bibnamefont
  {Tanigaki}}, \bibinfo {author} {\bibfnamefont {K.}~\bibnamefont {Takamiya}},
  \bibinfo {author} {\bibfnamefont {N.}~\bibnamefont {Sato}}, \bibinfo {author}
  {\bibfnamefont {R.}~\bibnamefont {Okumura}}, \bibinfo {author} {\bibfnamefont
  {Y.}~\bibnamefont {Uchihori}}, \ and\ \bibinfo {author} {\bibfnamefont
  {K.}~\bibnamefont {Saito}},\ }\href {\doibase 10.1016/j.jenvrad.2014.05.014}
  {\bibfield  {journal} {\bibinfo  {journal} {J. Environ. Radioactiv.}\
  }\textbf {\bibinfo {volume} {139}},\ \bibinfo {pages} {266} (\bibinfo {year}
  {2014})}\BibitemShut {NoStop}%
\bibitem [{\citenamefont {Tsuda}\ \emph {et~al.}(2014)\citenamefont {Tsuda},
  \citenamefont {Yoshida}, \citenamefont {Tsutsumi},\ and\ \citenamefont
  {Saito}}]{Tsuda2014}%
  \BibitemOpen
  \bibfield  {author} {\bibinfo {author} {\bibfnamefont {S.}~\bibnamefont
  {Tsuda}}, \bibinfo {author} {\bibfnamefont {T.}~\bibnamefont {Yoshida}},
  \bibinfo {author} {\bibfnamefont {M.}~\bibnamefont {Tsutsumi}}, \ and\
  \bibinfo {author} {\bibfnamefont {K.}~\bibnamefont {Saito}},\ }\href
  {\doibase 10.1016/j.jenvrad.2014.02.028} {\bibfield  {journal} {\bibinfo
  {journal} {J. Environ. Radioactiv.}\ }\textbf {\bibinfo {volume} {139}},\
  \bibinfo {pages} {260} (\bibinfo {year} {2014})}\BibitemShut {NoStop}%
\bibitem [{\citenamefont {Tanigaki}\ \emph {et~al.}(2015)\citenamefont
  {Tanigaki}, \citenamefont {Okumura}, \citenamefont {Takamiya}, \citenamefont
  {Sato}, \citenamefont {Yoshino}, \citenamefont {Yoshinaga}, \citenamefont
  {Kobayashi}, \citenamefont {Uehara},\ and\ \citenamefont
  {Yamana}}]{Tanigaki2015}%
  \BibitemOpen
  \bibfield  {author} {\bibinfo {author} {\bibfnamefont {M.}~\bibnamefont
  {Tanigaki}}, \bibinfo {author} {\bibfnamefont {R.}~\bibnamefont {Okumura}},
  \bibinfo {author} {\bibfnamefont {K.}~\bibnamefont {Takamiya}}, \bibinfo
  {author} {\bibfnamefont {N.}~\bibnamefont {Sato}}, \bibinfo {author}
  {\bibfnamefont {H.}~\bibnamefont {Yoshino}}, \bibinfo {author} {\bibfnamefont
  {H.}~\bibnamefont {Yoshinaga}}, \bibinfo {author} {\bibfnamefont
  {Y.}~\bibnamefont {Kobayashi}}, \bibinfo {author} {\bibfnamefont
  {A.}~\bibnamefont {Uehara}}, \ and\ \bibinfo {author} {\bibfnamefont
  {H.}~\bibnamefont {Yamana}},\ }\href {\doibase 10.1016/j.nima.2015.01.086}
  {\bibfield  {journal} {\bibinfo  {journal} {Nucl. Instrum. Methods Phys. Res.
  Sect. A}\ }\textbf {\bibinfo {volume} {781}},\ \bibinfo {pages} {57}
  (\bibinfo {year} {2015})}\BibitemShut {NoStop}%
\bibitem [{\citenamefont {Satoh}\ \emph {et~al.}(2014)\citenamefont {Satoh},
  \citenamefont {Kojima}, \citenamefont {Oizumi}, \citenamefont {Matsuda},
  \citenamefont {Iwamoto}, \citenamefont {Kugo}, \citenamefont {Sakamoto},
  \citenamefont {Endo},\ and\ \citenamefont {Okajima}}]{Satoh2014}%
  \BibitemOpen
  \bibfield  {author} {\bibinfo {author} {\bibfnamefont {D.}~\bibnamefont
  {Satoh}}, \bibinfo {author} {\bibfnamefont {K.}~\bibnamefont {Kojima}},
  \bibinfo {author} {\bibfnamefont {A.}~\bibnamefont {Oizumi}}, \bibinfo
  {author} {\bibfnamefont {N.}~\bibnamefont {Matsuda}}, \bibinfo {author}
  {\bibfnamefont {H.}~\bibnamefont {Iwamoto}}, \bibinfo {author} {\bibfnamefont
  {T.}~\bibnamefont {Kugo}}, \bibinfo {author} {\bibfnamefont {Y.}~\bibnamefont
  {Sakamoto}}, \bibinfo {author} {\bibfnamefont {A.}~\bibnamefont {Endo}}, \
  and\ \bibinfo {author} {\bibfnamefont {S.}~\bibnamefont {Okajima}},\ }\href
  {\doibase 10.1080/00223131.2014.886534} {\bibfield  {journal} {\bibinfo
  {journal} {J. Nucl. Sci. Technol.}\ }\textbf {\bibinfo {volume} {51}},\
  \bibinfo {pages} {656} (\bibinfo {year} {2014})}\BibitemShut {NoStop}%
\end{thebibliography}%

\end{document}